\documentclass[aps,twocolumn,pra]{revtex4}
\usepackage[dvips]{graphicx}
\begin{document}
\draft

\title{ Images of the Dark Soliton in a Depleted Condensate }

\author{ Jacek Dziarmaga, Zbyszek P. Karkuszewski, and Krzysztof Sacha}

\address{
Intytut Fizyki Uniwersytetu Jagiello\'nskiego, \\
ul.~Reymonta 4, 30-059 Krak\'ow, Poland 
}

\date{ December 07, 2002}

\begin{abstract}
The dark soliton created in a Bose-Einstein condensate becomes grey in 
course of time evolution because its notch fills up with depleted atoms. 
This is the result of quantum mechanical calculations which describes 
output of many experimental repetitions of creation of the stationary 
soliton, and its time evolution terminated by a destructive density 
measurement. However, such a description is not suitable to predict the 
outcome of a single realization of the experiment were two extreme 
scenarios and many combinations thereof are possible: one will see 
(1) a {\it displaced dark soliton} without any atoms in the notch, 
but with a randomly displaced position, or 
(2) a {\it grey soliton} with a fixed position, 
but a random number of atoms filling its notch. 
In either case the average over many realizations will reproduce the 
mentioned quantum mechanical result. In this paper we use $N$-particle 
wavefunctions, which follow from the number-conserving Bogoliubov theory, 
to settle this issue. 
\end{abstract}
\maketitle


\section{ Introduction }

  Let us start with a trivial but important remark on measurement in quantum
mechanics: in general a single measurement on an $N$-particle system is
neither equivalent to many measurements on the $N$-particle system, 
nor to $N$ measurements 
on $N$ one-particle systems. The only exception is the case of a product 
$N$-particle state where all particles are described by the same single-particle 
wavefunction. Not surprisingly the particles must be bosons forming a 
Bose-Einstein condensate, and a measurement on this $N$-particle state
can be replaced with $N$ measurements on $N$ one-particle systems in 
the same single-particle wavefunction.

   The importance of this "nonequivalence" principle can be seen in the
explanation of an interference experiment Ref.\cite{interf} in an excellent 
paper by Javanainen and Yoo \cite{JY}, and then elaborated in
Refs.\cite{localization}. In Ref.\cite{JY} the authors consider a Fock state
\begin{equation}
\left|\frac{N}{2},\frac{N}{2}\right\rangle \label{N/2}
\label{Fock}
\end{equation}
with $N/2$ atoms in a condensate of wavefunction $\phi_0(x)$ and $N/2$ in
a condensate with an orthogonal wavefunction $\phi_1(x)$. A single particle
density matrix of the state (\ref{N/2}) predicts that on average a density 
measurement will give a distribution of atoms
\begin{equation}
\frac{N}{2}\left(|\phi_0(x)|^2+|\phi_1(x)|^2\right)
\label{JYsingle}
\end{equation}
without any interference fringes between the two condensates. This single 
particle distribution is an average over many density measurements on the
same Fock state (\ref{N/2}). Simulation of a destructive 
density measurement \cite{JY}
based on the full $N$-particle wavefunction (\ref{N/2}), and the actual 
experiment \cite{interf}, give distributions with an interference term
\begin{eqnarray}
\frac{N}{2}&(&|\phi_0(x)|^2+|\phi_1(x)|^2+
\nonumber\\                 
&&
+\phi_0(x)\phi_1^*(x)e^{+i\varphi}+{\rm c.c.} )~.
\end{eqnarray}
For a given experimental realization the phase $\varphi$ is chosen randomly
from the interval $[0,2\pi)$. The outcome of a density measurement is the same 
as if before the measurement all $N$ atoms were not in the state (\ref{Fock}) 
but in an absurd single condensate with a wavefunction
\begin{equation}
\frac{1}{\sqrt{2}}\left(\phi_0+\phi_1e^{i\varphi}\right)~\label{JYvarphi}
\end{equation}
with a relative phase $\varphi$ that is not known before we actually
measure the density distribution. It turns out \cite{localization} that it
is enough to measure only a small fraction of the large total $N$ to prepare 
the remaining atoms in one of the condensate wavefunctions (\ref{JYvarphi}), or, 
in other words, to establish a definite phase $\varphi$. Destructive 
measurement of atomic positions, which effectively annihilates measured atoms 
from the trap, drives the remaining atoms into a condensate state with 
a randomly picked condensate wavefunction.
These interesting conclusions cannot be
obtained without full knowledge of the $N$-particle state (\ref{N/2}): the 
single particle density distribution (\ref{JYsingle}) is a (misleading) average 
over many experiments with different outcomes $\varphi$. To predict possible 
outcomes of individual measurements on an $N$-particle state it is essential to 
know its $N$-particle wavefunction.

  A perfect condensate is a state where all $N$ atoms are in the same single
particle state described by a single particle wavefunction $\phi_0(\vec x)$,
\begin{equation}
\left(\hat a_0^{\dagger}\right)^N|0\rangle~.
\end{equation}
When we include interactions between atoms but neglect depletion of atoms
from the condensate wavefunction, then $\phi_0$ is a solution of the celebrated 
Gross-Pitaevskii equation \cite{GPE}. Including small quantum depletion in the 
framework of the number-conserving Bogoliubov theory \cite{BT} leads to pair-
correlated eigenstates \cite{Z}. For example, the state without any 
quasiparticles (the Bogoliubov vacuum) has a pair-correlated form
\begin{equation}
(\hat d^\dagger)^{\frac{N}{2}}|0\rangle=
\left( \hat a_0^{\dagger} \hat a_0^{\dagger} +
       \sum_{k=1}^{\infty} \lambda_k \hat a_k^{\dagger} \hat a_k^{\dagger}
\right)^{\frac{N}{2}}|0\rangle~.\label{Z1}
\end{equation}
Operators $\hat a_k$ annihilate in a basis of functions $\phi_k$ orthogonal to 
the condensate wavefunction $\phi_0$. This general pair-correlated ansatz
(\ref{Z1}) has been known for a while, see e.g. the review by Leggett 
\cite{Leggett}, but a general solution for the coefficients $\lambda_k$ and 
corresponding wavefunctions $\phi_k$ has been found only very recently in 
Ref.\cite{Z} (the solution in the special case of a uniform condensate was given 
in Ref.\cite{Leggett}). The two-particle creation operator $d^\dagger$ has to
commute with all quasiparticle annihilation operators (see Appendix A)
\begin{equation}
[\hat b_m, d^\dagger]=0, \; \mbox{for all}\; m.
\end{equation}
The $N$-particle state (\ref{Z1}) is a foundation on which one can build 
theory of BEC entirely in the language of $N$-particle wavefunctions.

  In our two recent papers \cite{greyI,greyII} we studied depletion from a 
condensate with a dark soliton. The dark soliton state is a collectively
excited condensate with a notch in the condensate wavefunction. In the
Thomas-Fermi limit of strong interactions the notch behaves like a soliton
with a wavefunction close to the notch proportional to
\begin{equation}
\phi_0(x)\sim\tanh(x-X)~,
\end{equation}
where $X$ is a position of the soliton measured in units of the healing length. 
The anomalous Bogoliubov mode with a 
wavefunction 
\begin{equation}
\phi_1(x)\sim -\partial_X\phi_0(x)=\cosh^{-2}(x-X) 
\label{modeX}
\end{equation}
dominates quantum depletion from the condensate within the soliton. In 
Ref.\cite{greyII} we found an average number of atoms $dN$ depleted into
the non-condensate mode $\phi_1$ and a single particle density distribution
which close to the soliton at $X=0$ looks like
\begin{eqnarray}
p_1(x)&=&
(N-dN) |\phi_0(x)|^2 +
dN   |\phi_1(x)|^2 ~.
\end{eqnarray}
For $dN=0$ the distribution has a hole at $X=0$, but with increasing quantum 
depletion $dN$ the hole is filling with atoms. This single particle result means 
that on average over many experiments the notch of a depleted dark soliton
appears filled with atoms (it appears grey, not dark), but it gives no clue what 
to expect as an outcome of a single destructive density measurement. We address 
this issue in the following sections.
 
   The paper is organized as follows. In Sections 
\ref{darkTF},\ref{darkND},\ref{darkGNS} we apply the number conserving 
Bogoliubov theory (shortly summarized in Appendix A) to the dark soliton, and 
simulate density measurement on different $N$-particle quantum states that all 
may pass under the same label: ``depleted dark soliton''. We find that images of 
the ``depleted dark soliton'' depend qualitatively on the actual $N$-particle 
state. To simulate the measurements we adapted the numerical algorithm of 
Ref.\cite{JY} which is described in Appendix B. We conclude in Section 
\ref{conclusion}.

\section{ Stationary Dark Soliton }
\label{darkTF}

   The dark soliton state \cite{DS} of a condensate in a 1D harmonic trap can be 
defined as an antisymmetric solution of the stationary Gross-Pitaevskii equation
\begin{equation} 
- \frac12 \partial_x^2 \phi_0 + 
\frac12 x^2 \phi_0 + 
g |\phi_0|^2 \phi_0=\mu\phi_0 \; 
\label{GP} 
\end{equation}
In this paper we use the dimensionless oscillator units. In the Thomas-Fermi
(TF) limit of $g\gg 1$ the dark soliton state is
\begin{equation}
\phi_0(x) \; \approx \;
F(x)\; \tanh\left(x/l_0\right) ,
\label{phi0}
\end{equation}
with
\begin{equation}
F(x)=\sqrt{ \frac{2\mu-x^2}{2g} } ,
\label{TF}
\end{equation}
where the chemical potential $\mu\approx(3g/2)^{2/3}/2$ and the healing length 
$l_0\approx\sqrt{2}\left(2/3g\right)^{1/3}$.

  A perfect condensate with all $N$ atoms in the dark soliton mode $\phi_0$  
is not an energy eigenstate because interactions between atoms continuously 
deplete atoms from the condensate wavefunction $\phi_0$. The Bogoliubov
theory is a systematic way to describe small quantum fluctuations around 
the Gross-Pitaevskii solution, and to find stationary states of a depleted
condensate. 

  Solution of the Bogoliubov-de Gennes (BdG) equations (see Appendix A) reveals 
many different modes of the quasiparticle spectrum. However, only the negative 
energy (anomalous) mode (\ref{modeX}) 
has a wavefunction localized in the soliton notch \cite{Law,greyII,Z}. 
>From all modes the anomalous mode will contribute the most to the density of 
incoherent atoms in the soliton notch. In the following, similarly as in 
Ref.~\cite{greyII}, we truncate the Bogoliubov spectrum to the anomalous mode 
alone and look for the Bogoliubov vacuum state in the form
\begin{equation}
|0_b:N\rangle \sim (\hat a_0^\dagger \hat a_0^\dagger+\lambda\hat a_1^\dagger
\hat a_1^\dagger)^{\frac{N}{2}}\;|0 \rangle
\label{ZDS}
\end{equation}
with $\hat a_1^\dagger$ creating in the mode $\phi_1$.
  The anomalous mode solution of the BdG equations (\ref{BdG}) with 
$\omega=\frac{-1}{\sqrt{2}}$ is
\begin{equation}
u_1=\frac12\left( f_+ + f_- \right)~,~~
v_1=\frac12\left( f_+ - f_- \right)~,
\label{uvS}
\end{equation}
with functions
\begin{eqnarray}
f_+(x)=\frac{\sqrt{3g}}{\sqrt{8}\cosh^2\left(\frac{x}{l_0}\right)}~,~~
f_-(x)=\sqrt{\frac23}F(x)~, 
\label{f+-}
\end{eqnarray}
see Ref.\cite{greyII}. For $g\gg 1$ equations (\ref{uvS}) are dominated by 
$f_+$, $\langle f_+|f_+ \rangle\gg\langle f_-|f_- \rangle$, and the two 
functions become the same, $u_1\approx v_1\approx\frac12 f_+$. As a result in 
the TF limit the anomalous mode does not mix with other modes and the dominant 
eigenfunction is 
\begin{equation}
\phi_1(x)=\frac{f_+(x)}{\sqrt{\langle f_+|f_+ \rangle}}~,
\end{equation}
as has been verified numerically in Ref.\cite{Z}. To find the asymptotic 
behavior of $\lambda$ for $g\gg 1$ we have to be more careful and still keep 
$f_-$ for a moment. The eigenvalue can be obtained from the general prescription 
presented in Appendix~A and it is
\begin{equation}
\lambda\approx \frac{\langle f_+|f_+ \rangle-\langle f_+|f_- \rangle}
{\langle f_+|f_+ \rangle+\langle f_+|f_- \rangle} \approx 
1-\frac{2^{7/6}3^{1/3}}{g^{2/3}}~.\label{lambdaS}
\end{equation}
  Knowing $\lambda$ we can find an average number of depleted atoms $dN$ 
in the Bogoliubov vacuum state (\ref{ZDS}). In the particle Fock representation 
the unnormalized vacuum state (\ref{ZDS}) is a sum
\begin{eqnarray}
|0_b:N\rangle & \sim &
\sum_{k=0}^{\frac{N}{2}}
\lambda^k          
\frac{\sqrt{    (N-2k)! \; (2k)! }}
           { (N/2-k)! \; k! }
|N-2k,2k\rangle.
\label{solZ} 
\end{eqnarray} 
For depletion greater than unity but small with respect to the total number of
atoms $N$, i.e. $1\ll dN \ll N$, we may use the Stirling formula, and then 
replace the sum by an integral. As a result we find expressions for $dN$ and its 
dispersion $D(dN)$
\begin{eqnarray}
&& dN\approx
-\frac{1}{\ln\lambda}\approx\frac{g^{2/3}}{4~2^{1/6}~3^{1/3}} ~, \label{dN0}\\
&& D(dN)=dN.  \label{DdN0}
\end{eqnarray}

  The single particle density distribution in the vacuum state 
(\ref{ZDS},\ref{solZ}) is
\begin{equation}
p_1(x)=(N-dN)|\phi_0(x)|^2+dN|\phi_1(x)|^2.
\label{singSS}
\end{equation}
The first term on the RHS is the dark soliton density profile with a hole at 
$x=0$, while the second one stands for the depleted atoms localized in the 
soliton notch. A good measure of depletion is a ratio of atomic density
in the notch at $x=0$ to the density near the notch
\begin{equation}
\frac{dN|\phi_1(0)|^2}{NF(0)^2} \sim N^{1/3}~,
\label{Sscal}
\end{equation}
suggesting that for sufficiently large number of atoms, the notch will not
be visible. For the parameters of the Hannover experiment \cite{Hannover},
where $g\approx 7500$ and $N=1.5\times10^5$, $dN\approx 60$ atoms and 
the ratio is $15\%$.
 
   However, Eq.~(\ref{singSS}) predicts an average over many experimental 
realizations. For a single experiment one has to choose randomly positions of 
$N$ atoms according to the $N$-particle probability distribution. In general 
this is rather a formidable task, but, as shown by Javanainen and Yoo \cite{JY} 
(see Appendix~B), for a state spanned on only two modes, and
with a sequential algorithm that chooses a position of $n+1$-st atom after
positions of $n$ atoms have already been ``measured'', this problem is 
polynomial in $N$. 

  In general we expect that a single measurement of a depleted dark soliton
will show the same atomic density distribution as if all $N$ atoms were 
initially prepared in a condensate with one of wavefunctions  
\begin{equation}
\phi_0(x)-\phi_1(x) q e^{i\varphi}~.
\label{qvarphi}
\end{equation}
The real parameters $(q,\varphi)$ are chosen at random and fluctuate from
experiment to experiment. 
In the special case of $\varphi=0,\pi$  
the condensate wavefunction (\ref{qvarphi}) is simply a {\it displaced dark 
soliton} 
\begin{equation}
\phi_0(x)-\phi_1(x) q\approx 
F(x)\tanh\left(\frac{x-q\sqrt{\frac{3l_0}{4}}}{l_0}\right)~.
\end{equation}
In general, when $\varphi\neq 0,\pi$, the condensate wavefunction 
(\ref{qvarphi}) gives a density distribution 
\begin{eqnarray}  
&& N|\phi_0-\phi_1 q e^{i\varphi}|^2=
\nonumber\\
&& N\left[\phi_0-\phi_1 q \cos \varphi\right]^2+
   N\phi_1^2q^2\sin^2\varphi~.     
\label{qphi}             
\end{eqnarray}
For small $q$ this distribution can be interpreted as a soliton shifted to 
$q\sqrt{\frac{3l_0}{4}}\cos\varphi$, but with a nonzero atomic density  
$N\phi_1^2q^2\sin^2\varphi$ in the notch. For $\varphi\neq 0,\pi$ the soliton 
appears as a {\it displaced grey soliton}.
\begin{figure}[htb]
\includegraphics*[width=8.6cm]{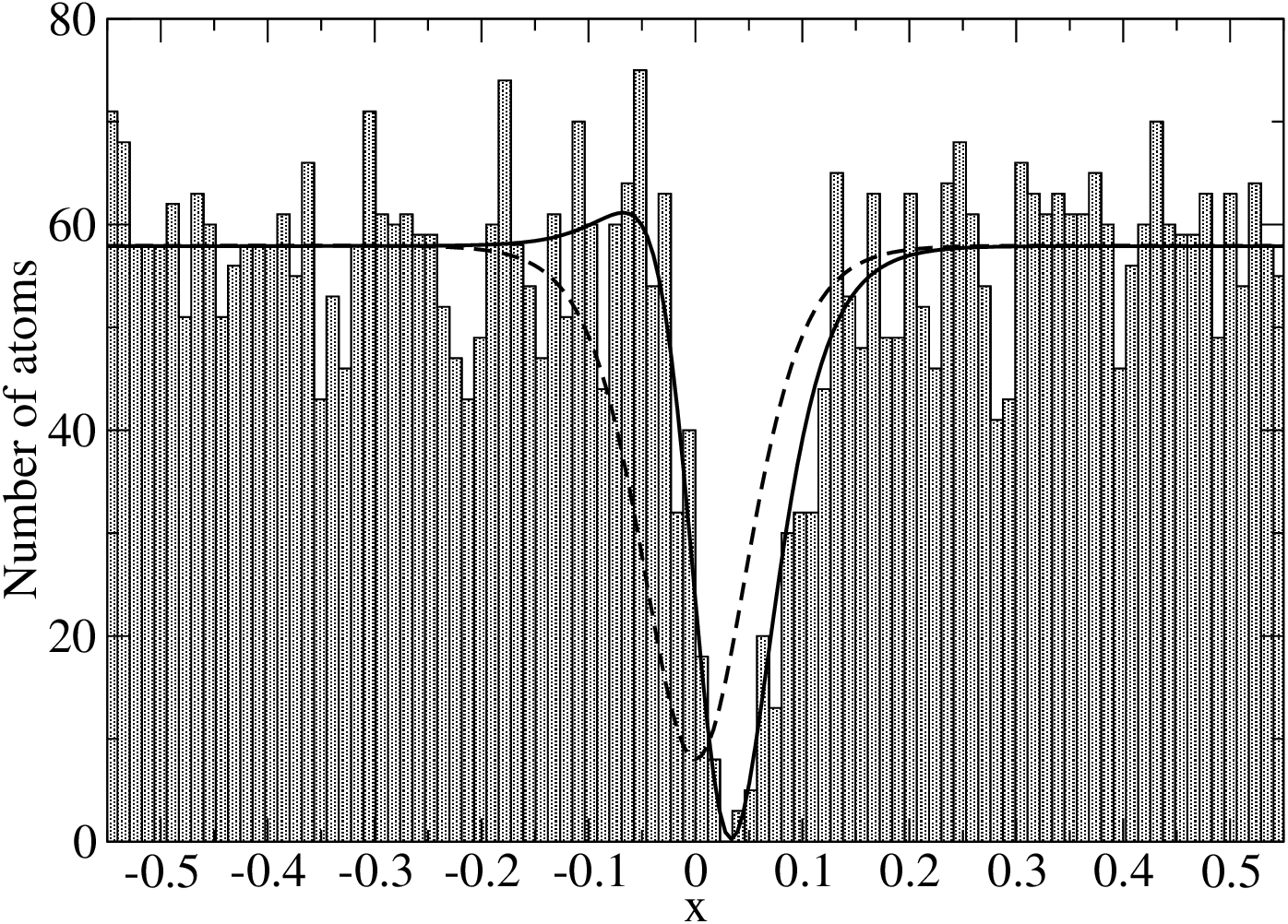}
\caption{Density measurement for state $|0_b:N\rangle$,  
Eq.{\protect (\ref{solZ})}. The histogram represents
number of particles in bins around the solitonic notch measured 
in a single experiment. The solid line
stands for $\bar p(x)=\bar A |\phi_0(x)|^2+\bar B |\phi_1(x)|^2 + 
\bar C\phi_0(x)\phi_1(x)$, where $\bar A, \bar B, \bar C$ are parameters
of {\protect (\ref{MarProb})} averaged over all atoms, see the formula 
{\protect (\ref{MarProb})}. The dashed line gives the expected
average outcome of many experiments i.e. the single particle density.
The values of the parameters correspond to 
the Hannover experiment \cite{Hannover}, where $g=7500$ and
$N=1.5\times10^5$. } 
\label{fig1}
\end{figure}

\begin{figure}[htb]
\includegraphics*[width=8.6cm]{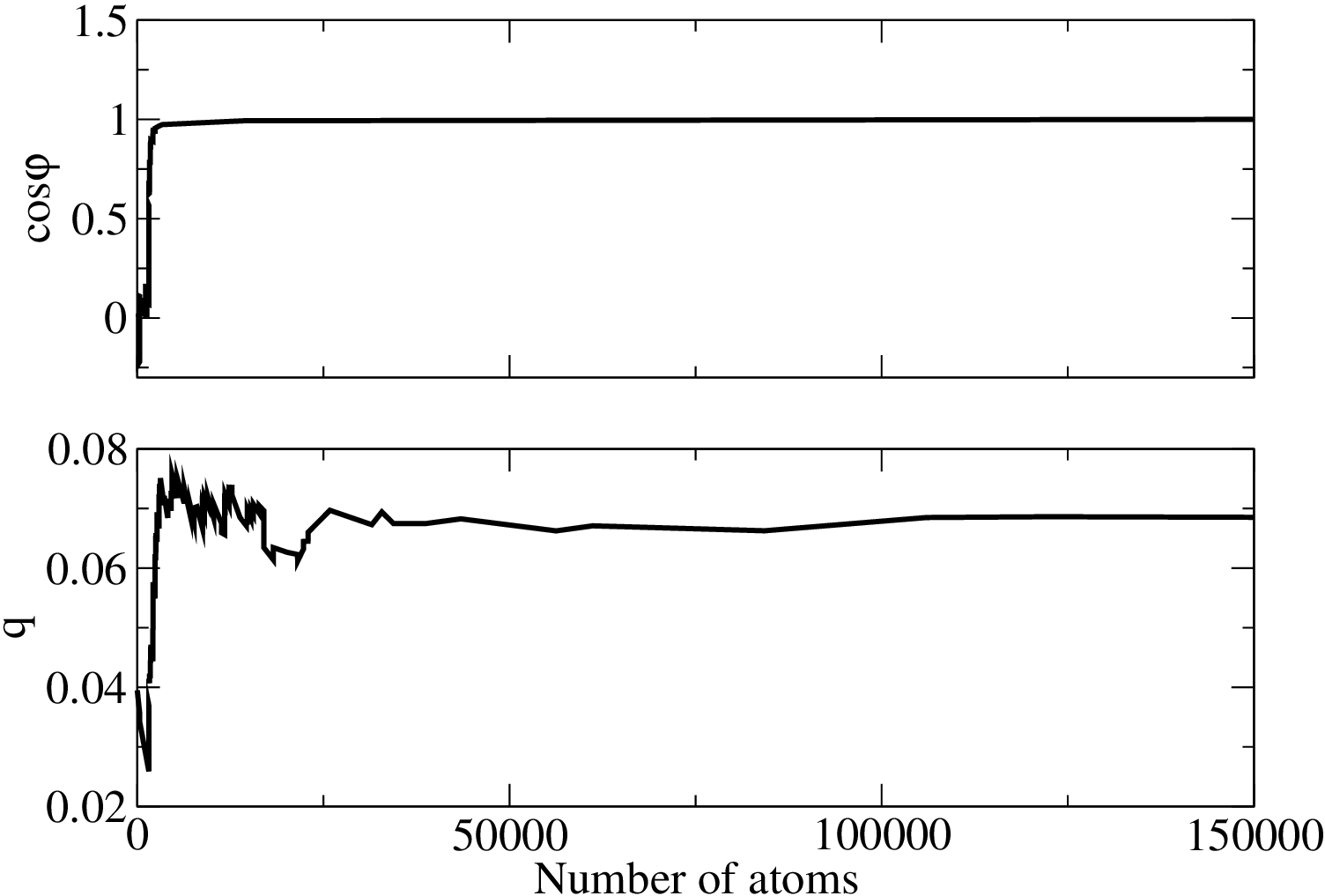}
\caption{Values of $\cos\varphi$ (upper plot) and $q$ during the density 
measurement performed on the state $|0_b:N\rangle$.} 
\label{fig2}
\end{figure}

  Adopting the scheme of Javanainen and Yoo we generated 
results of individual density measurements of the Bogoliubov vacuum state 
(\ref{ZDS},\ref{solZ}). A generic outcome is a displaced dark soliton shown in 
Fig.~\ref{fig1}. In Fig.~\ref{fig2} we show how definite values of 
$\cos(\varphi)$ 
and $q$ are established in the course of measurement of subsequent atomic 
positions. It is enough to measure only a small fraction of atoms to suppress 
fluctuations of $\varphi$ and $q$, see Fig.~\ref{fig2}.

  A density measurement of the Bogoliubov vacuum state (\ref{ZDS},\ref{solZ})
always gives a displaced dark soliton with $\varphi=0$ in Eq.(\ref{qvarphi}). 
At first sight this may seem obvious because the wavefunctions
$\phi_0$ and $\phi_1$ are real, and all amplitudes in the state (\ref{solZ})
are also real. However, this simple argument does not take into account 
the possibility that the measurement of a small fraction of atoms may localize
the state of remaining atoms not in a single condensate (\ref{qvarphi}), but
in a superposition of several condensates \cite{localization} with different 
parameters $(q,\varphi)$. Indeed, as we can see in Eq.(\ref{qphi}), density 
distributions for $(q,+\varphi)$ and $(q,-\varphi)$ are the same so the density 
measurement cannot distinguish $+\varphi$ from $-\varphi$. A real superposition 
of two condensates with opposite phases has real amplitudes, just like the state 
(\ref{solZ}). We checked that a measurement of a Fock state $|N-n,n\rangle$, 
instead of the Bogoliubov vacuum state (\ref{ZDS},\ref{solZ}), 
indeed results in a real superposition of two condensates with nonzero 
$+\varphi$ and $-\varphi$, which appears as a displaced grey soliton. What makes
a difference is the fact the Bogoliubov vacuum (\ref{solZ}) is a superposition 
over many Fock states. It is the phase coherence between different Fock states, 
all with real amplitudes, that enforces the observed $\varphi=0$.  
  
\section{ Soliton Condensate without Initial Depletion }
\label{darkND}

  In current experiments \cite{Hannover,DS2} the dark solitons are generated 
with the help of the phase imprinting \cite{PIM}. As we argued in 
Ref.~\cite{greyII}, when the depletion in the condensate before the imprinting
is negligible, then the soliton state right after imprinting can be idealized by 
the state without any atoms depleted from the solitonic condensate $\phi_0$ to 
the anomalous mode $\phi_1$,
\begin{equation}
|\Psi(t=0)\rangle=|N,0\rangle~.
\label{initPS}
\end{equation}
This state is not a stationary state. To obtain time evolution of this state we 
consequently truncate the Bogoliubov Hamiltonian to the anomalous mode where
\begin{eqnarray}
H&=&-\frac{1}{\sqrt{2}}\hat b_1^{\dagger}\hat b_1   
\end{eqnarray}
with
\begin{eqnarray}
\hat b_1&=&
\frac{\hat a_0^{\dagger}\hat a_1-\lambda\hat a_0\hat a_1^{\dagger}}
     {\sqrt{N(1-\lambda^2)}}~.
\end{eqnarray}
With this Hamiltonian the initial perfect condensate 
$|\Psi(t=0)\rangle$ evolves 
\begin{figure}[htb]
\includegraphics*[width=8.6cm]{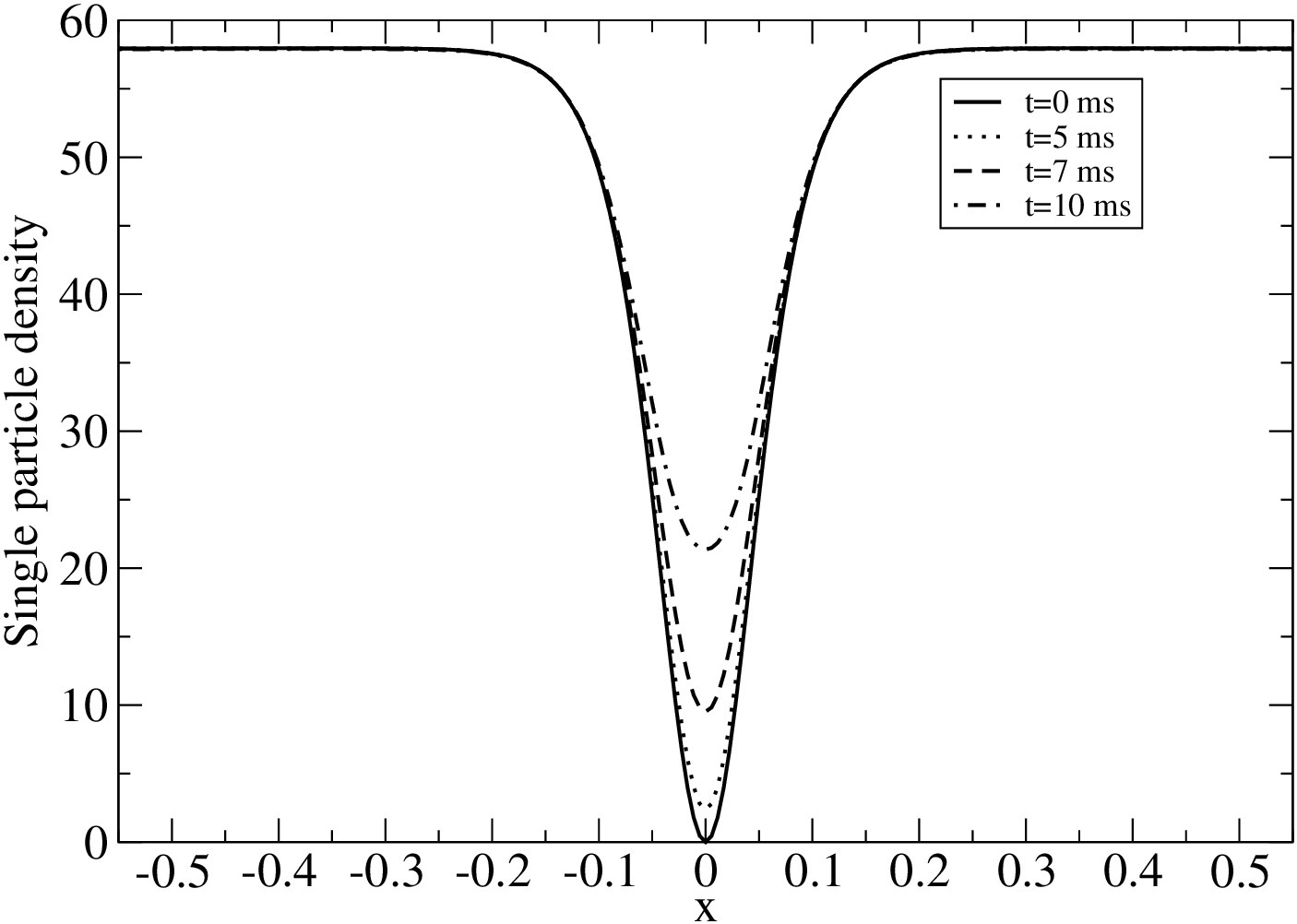}
\caption{Single particle densities of the state $|\Psi(t)\rangle$, Eq.{\protect 
(\ref{initPS})}, at different 
instants of time.
The values of the parameters correspond to 
the Hannover experiment \cite{Hannover}, where $g\approx 7500$ and
$N=1.5\times10^5$.}
\label{singPS}
\end{figure}
into a depleted state $|\Psi(t)\rangle$. In Fig.~\ref{singPS} we show how a 
single particle density evolves in time. 
After about 15~ms (for the 
parameters of the Hannover experiment \cite{Hannover}) the hole in the single
particle density distribution fills up with depleted atoms.

\begin{figure}[htb]
\includegraphics*[width=8.6cm]{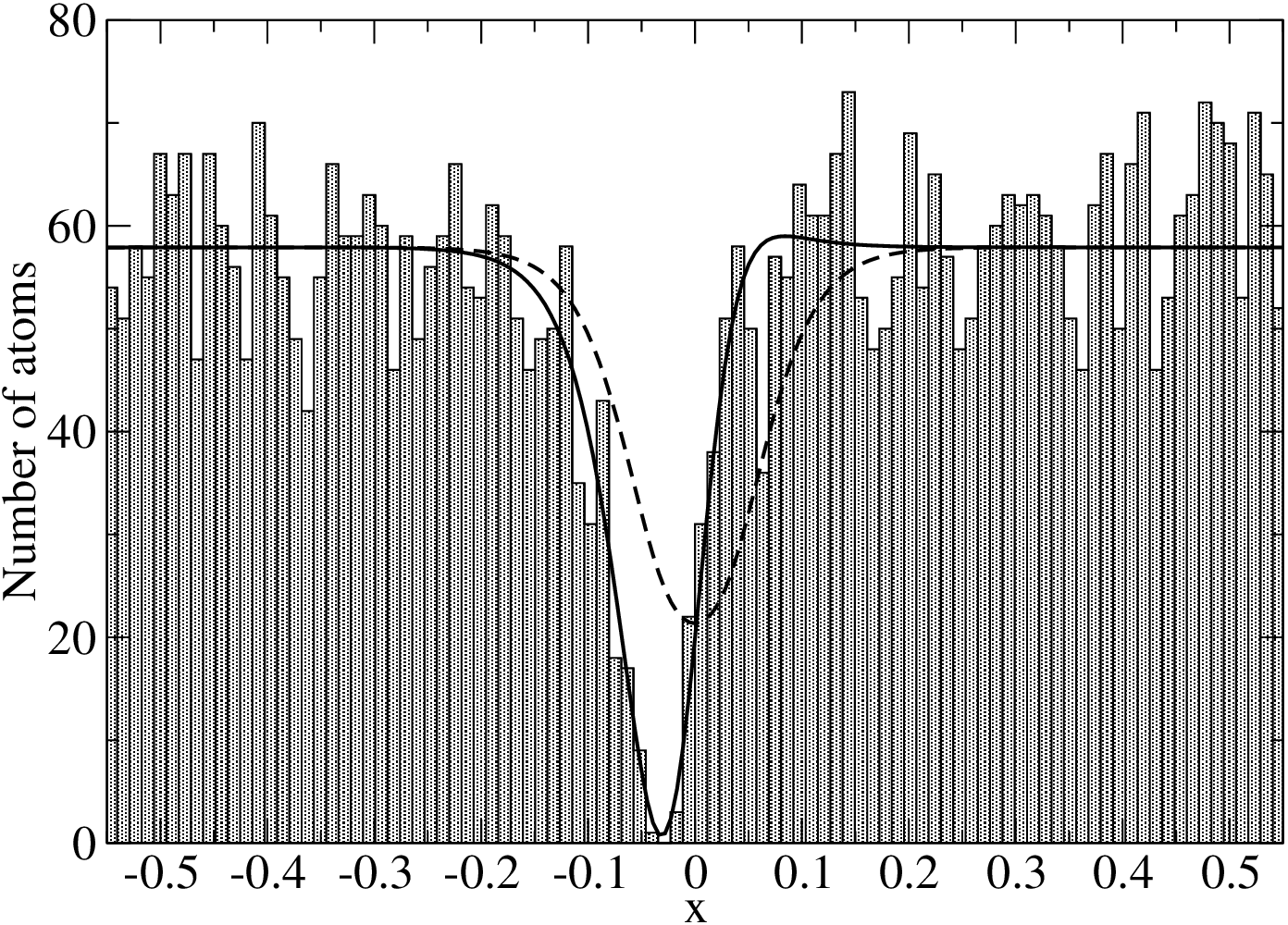}
\caption{The same as Fig.\ref{fig1} but for the state $|\Psi(t=0.15)\rangle$.} 
\label{histPS}
\end{figure}

  We simulated single realizations of destructive density measurements of 
 the state 
$|\Psi(t)\rangle$. A generic density distribution after evolution for 
10~ms is shown in Fig.~\ref{histPS}. We see that the soliton is dark and it is 
displaced with respect to the trap center. 
  Inspection of the state after 10~ms shows that
$\langle N-k,k|\Psi(10~{\rm ms})\rangle \sim e^{ik\varphi}$ with a tiny
$\varphi\approx 0.01$. In principle the time evolution might generate
any phase $\varphi\in[0,2\pi)$, but the parameters of the experiment 
\cite{Hannover} are such that there is not enough time to get any more 
significant $\varphi$ from the idealized initial state (\ref{initPS}) before 
the linearized Bogoliubov theory breaks down. 

\section{ Generic Non-Stationary Soliton States }
\label{darkGNS}

  In a realistic experiment the dark soliton is created in neither the 
stationary nor the zero depletion state discussed in the previous two 
Sections. The two states may appear ``natural'' to a theorist's eye 
accustomed to such highly idealized examples. However, these states are
not realistic from experimental point of view. The soliton is phase
imprinted on a condensate with a $10\%$ 
thermal cloud i.e. in a highly 
excited non-stationary state. We do not know what the actual state, or ensemble 
of states, is, but we can still test how robust is the picture developed in the
previous Sections with respect to imperfections of the quantum state. 
 
\begin{figure}[htb]
\includegraphics*[width=8.6cm]{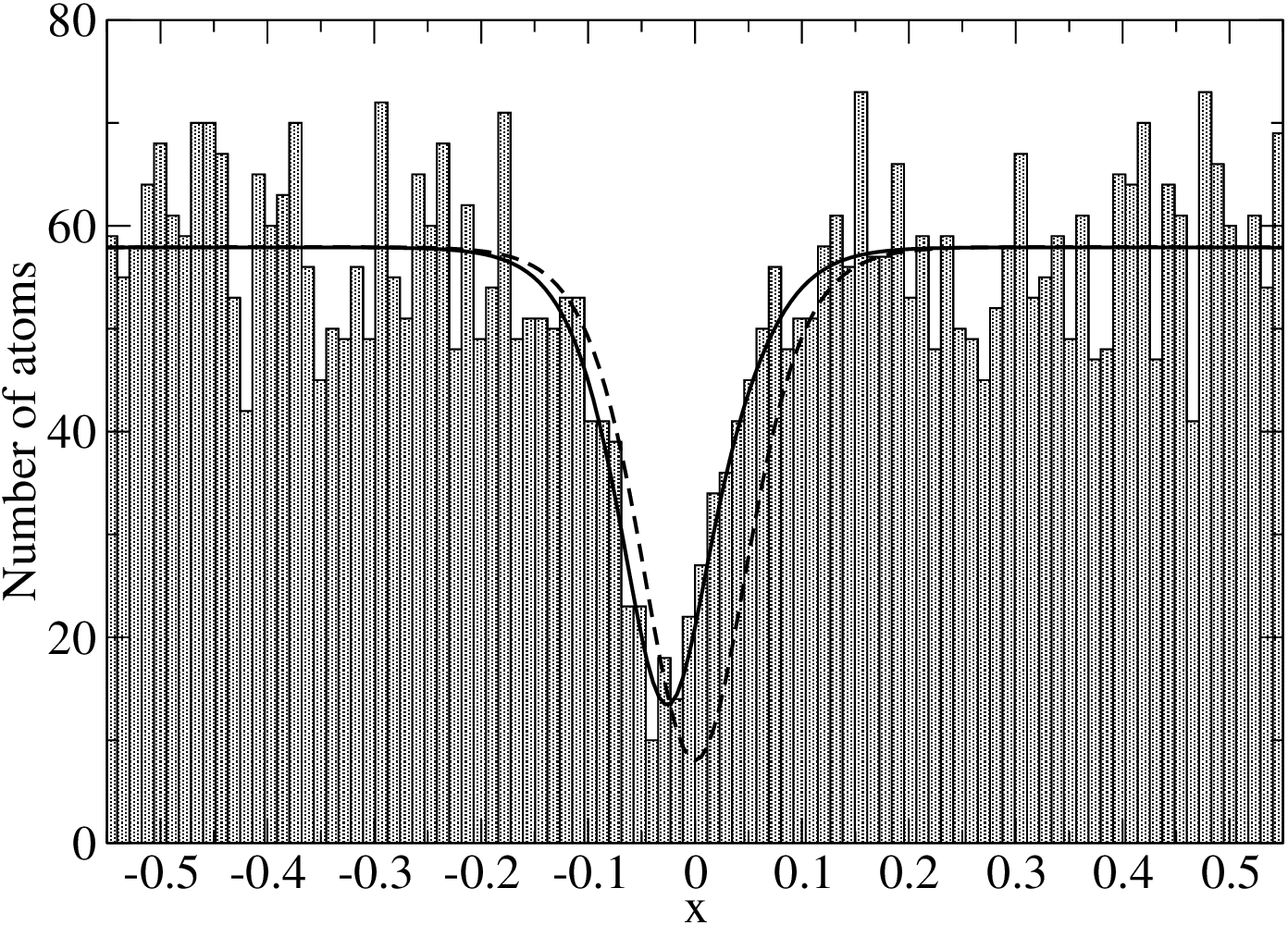}
\caption{The same as Fig.\ref{fig1} but for the state $|{\rm
rand}\rangle$.} 
\label{randS}
\end{figure}

\begin{figure}[htb]
\includegraphics*[width=8.6cm]{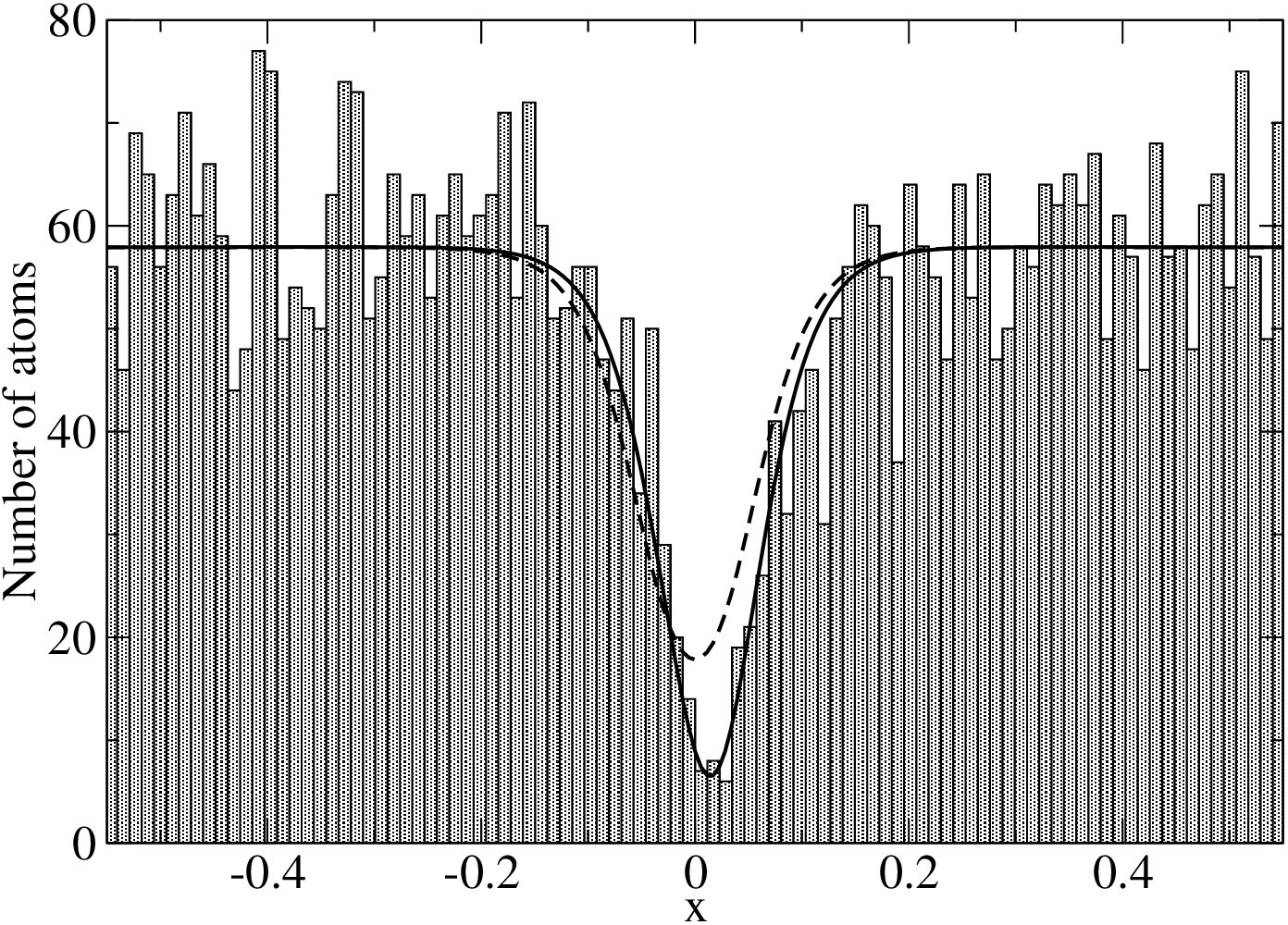}
\caption{The same as Fig.\ref{fig1} but for the state $|{\rm rand}\rangle$
after evolution over 0.7~ms.} 
\label{randt1}
\end{figure}

  To get an idea what are images of a generic non-stationary state we model
the state by a state with random phases. The Hamiltonian of the system is
diagonal in the Fock basis of {\it quasiparticles} so we expect a generic 
non-stationary state in to have apparently random phases in {\it particle} 
Fock representation. To be more specific, we take as an example the 
Bogoliubov vacuum (\ref{solZ}) with randomized phases $\alpha_k$,  
\begin{equation}
|{\rm rand}\rangle \sim
\sum_{k=0}^{N/2}
e^{i\alpha_k} \lambda^k          
\frac{\sqrt{    (N-2k)!   (2k)! }}
           { (N/2-k)!k! }
|N-2k,2k\rangle~. 
\label{0alphaPsi}
\end{equation}
Single particle density in this state is exactly the same as for the 
Bogoliubov vacuum, see Eq.~(\ref{singSS}), but possible outcomes 
of density measurements can be very different in the two states.
Indeed, a generic outcome in the state $|{\rm rand}\rangle$ shown in 
Fig.~\ref{randS} reveals a displaced grey soliton with 
$\cos\varphi=0.45$. The phase disorder present in the state 
$|{\rm rand}\rangle$ translates into random $\varphi$'s.
  As a second example, we take the state $|{\rm rand}\rangle$ as an
initial state instead of the state (\ref{initPS}), evolve it for
0.7~ms, and then measure atomic positions. In Fig.\ref{randt1}
we show a single particle density after 0.7~ms and a corresponding
generic outcome of density measurement. Not surprisingly,
we again see a displaced grey soliton, this time with $\cos\varphi=-0.48$.  

  These two examples demonstrate that phase disorder in particle Fock 
representation can change qualitatively possible results of density 
measurements. 

\section{ Conclusion }
\label{conclusion}

In Refs.\cite{greyI,greyII} we have considered influence of quantum
fluctuations on a density profile of a Bose-Einstein condensate excited
to a solitonic state. We have shown that the soliton notch may be filled with 
atoms that are depleted from the condensate due to interactions between
particles. The analysis has been performed by calculating a single particle
density matrix. However, such a matrix provides informations about
the density profile that is a result of average over many experimental 
realizations. To obtain predictions for a single experimental realization
one has to choose randomly positions of atoms according to the multiparticle
probability density. 

In this paper we concentrate on the analysis of outcome of the 
destructive density
measurement in a single experiment. 
Applying the number conserving Bogoliubov theory of a BEC we
calculate $N$-particle quantum states of a depleted condensate with a dark 
soliton. The $N$-particle wavefunctions are used to see what happens after 
the depleted condensate is subject to an ideal destructive (density) 
measurement of atomic positions.  
We have found that:

\begin{itemize}

\item 
For a BEC excited to a stationary state, i.e. the Bogoliubov vacuum state
without 
any quasiparticles, a destructive measurement reveals density distribution 
which appears like a displaced {\bf dark} soliton with an empty notch. Average 
displacement is proportional to the average number of incoherent atoms in the 
notch.

\item
For the non-stationary state that evolves from an initial 
state without any incoherent atoms in the soliton notch (i.e. a 
perfect condensate
with a dark soliton) a destructive measurement reveals, similarly like in the 
previous case, a displaced {\bf dark} soliton. 
The longer the evolution lasts the larger the depletion of the condensate
becomes (and thus also average displacement of the soliton).

\item 
The previous two cases are related to highly idealized states of a BEC.
To analyze influence of imperfections (in the preparation of the soliton
condensate) on the result of density measurement we introduced random phases 
in amplitudes of different Fock states corresponding to the Bogoliubov vacuum
state. In such a non-stationary state the measurement reveals a displaced 
{\bf grey} 
soliton. Both the average displacement and the average density of atoms in 
the notch of this {\bf grey} soliton 
are proportional to the number of atoms depleted from the condensate.

\end{itemize}

Very different $N$-particle quantum states may correspond to the same single
particle density matrix. However, different states result in qualitatively 
different results of destructive density measurements. 
In view of the strong dependence of measurement results on a quantum state,
it is an open question what $N$-particle state is relevant for the present day 
dark soliton experiments \cite{Hannover}. The phase imprinting 
method \cite{PIM} prepares neither an energy eigenstate nor the initial state 
without any depleted atoms. Certainly, further experiments would be helpful 
to settle this problem.

  Quantum fluctuations of a condensate are an intriguing result of 
interatomic interactions that introduce two-particle and higher correlations
 into the state of the system. BEC as a correlated system might be 
a very interesting toy model for condensed matter physics.
Depletion to the notch of the dark soliton gives a unique 
opportunity to detect quantum fluctuations simply by density measurement. 
A dark soliton in a strictly one-dimensional trap is well within the reach 
of present technology.


\section*{ APPENDIX }


\subsection{The N-particle State of BEC }
\label{StateZ}

  Here we briefly summarize main conclusions of the number-conserving
Bogoliubov theory of BEC \cite{BT,Z}. A field operator $\hat\psi(\vec x)$
is split into a condensate part $\hat a_0 \phi_0(\vec x)$ and a
non-condensate part $\delta\hat\psi(\vec x)$,
\begin{equation}
\hat\psi(\vec x)=\phi_0(\vec x)\hat a_0+\delta\hat\psi(\vec x)~~.
\end{equation}
To zero order in the ``small" fluctuation operator $\delta\hat\psi(\vec x)$
the $N$-particle state is a perfect condensate
\begin{equation}
\left( \hat a_0^{\dagger} \right)^N~|0\rangle
\end{equation}
with all $N$ atoms in the condensate wavefunction $\phi_0(\vec x)$.
The $\phi_0$ solves a stationary Gross-Pitaevskii equation (GPE)
\begin{equation}
\mu \phi_0=-\frac12 \nabla^2\phi_0 + V(\vec x)\phi_0 + g|\phi_0^2|\phi_0~.
\label{GPE}
\end{equation}
To zero order the interaction affects only the shape of the condensate 
wavefunction $\phi_0(\vec x)$ through the nonlinear term $g|\phi_0^2|\phi_0$
in the GPE. 

   Quantum depletion from this condensate shows up in the second order of the 
perturbation theory where the $N$-particle state becomes a pair-correlated 
Bogoliubov vacuum state
\begin{equation}
|0_b:N\rangle\sim
\left(
d^{\dagger}
\right)^{\frac{N}{2}}|0\rangle =
\left( \hat a_0^{\dagger} \hat a_0^{\dagger} +
       \sum_{k=1}^{\infty} \lambda_k \hat a_k^{\dagger} \hat a_k^{\dagger}
\right)^{\frac{N}{2}}|0\rangle.
\label{Z}
\end{equation}
The sum runs over an orthonormal basis of non-condensate modes
$\phi_k(\vec x)$ orthogonal to $\phi_0(\vec x)$. The eigenvalues
$\lambda_k$ and the eigenmodes $\phi_k(\vec x)$ have been calculated in
full generality only very recently \cite{Z}. Their construction runs in
the following steps. Its full justification can be found in
Refs.\cite{BT,Z}.

\begin{itemize} 

\item Solve the stationary GPE (\ref{GPE}) to get $\phi_0(\vec x)$ 
and $\mu$. Note that $\phi_0$ does not need to be the ground state, it can be 
an excited stationary state with a dark soliton or vortex.

\item Solve the linear Bogoliubov-de Gennes equations for the Bogoliubov normal 
modes of small fluctuations around $\phi_0$,
\begin{eqnarray}
&&
-\frac12 \nabla^2 U_m +
V(\vec x) U_m +
2g|\phi_0|^2 U_m +
g\phi_0^2 V_m =          \nonumber\\
&&
\mu U_m+\omega_m U_m~,   \nonumber\\
&&
-\frac12 \nabla^2 V_m +
V(\vec x) V_m +
2g |\phi_0|^2 V_m +
g(\phi_0^*)^2 U_m =      \nonumber\\
&&
\mu V_m-\omega_m V_m     \label{BdG}
\end{eqnarray} 
to get the eigenvalues $\omega_m$ and ``raw" Bogoliubov eigenmodes 
$U_m(\vec x),V_m(\vec x)$.

\item Project the raw $U_m,V_m$ on the subspace orthogonal to $\phi_0$,
\begin{eqnarray}
u_m &=& U_m - \phi_0 \langle \phi_0 | U_m \rangle~, \nonumber\\ 
v_m &=& V_m - \phi_0 \langle \phi_0 | V_m \rangle~,
\end{eqnarray}  
to get the Bogoliubov eigenmodes $(u_m,v_m)$.

\item Normalize the modes $(u_m,v_m)$ so that a norm
\begin{equation}
\langle u_m | u_m \rangle - \langle v_m | v_m \rangle = +1 ~.
\end{equation}
Ignore modes with negative norm.

\end{itemize}

  The normalized Bogoliubov modes define bosonic quasiparticle annihilation
operators
\begin{equation}
\hat b_m =
\frac{\hat a_0^{\dagger}}{\sqrt{N}} \langle u_m | \delta\hat\psi \rangle -
\frac{\hat a_0}{\sqrt{N}} 
\left( \langle v_m^* | \delta\hat\psi  \rangle \right)^{\dagger}~
\end{equation}
which transfer atoms between the condensate and the non-condensate modes,
but conserve the total number of atoms $N$. For small depletion the 
Hamiltonian can be approximated by a sum of harmonic oscillators
\begin{equation}
\hat H ~=~ \sum_m ~\omega_m~ \hat b_m^{\dagger}\hat b_m~.
\end{equation}
The $N$-particle Bogoliubov vacuum (\ref{Z}) is the eigenstate of $\hat H$ 
without any quasiparticles,
\begin{equation}
\hat b_m |0_b:N\rangle=0~.
\end{equation}
The $N$-particle state (\ref{Z}) is annihilated by all $\hat b$'s when the 
$\hat d^{\dagger}$ defined in Eq.(\ref{Z}) commutes with all $\hat b$'s,
\begin{equation}
\left[\hat b_m,\hat d^{\dagger}\right]=0~.
\end{equation}
In general construction of such a $\hat d^{\dagger}$ runs along the following 
steps. 

\begin{itemize}

\item Choose an orthonormal basis of states $\tilde\phi_k(\vec x)$ orthogonal to 
$\phi_0$, and then calculate matrices $\tilde U$ and $\tilde V$ with elements 
\begin{eqnarray}
&& \tilde U_{mk} = \langle u_m  |  \tilde\phi_k   \rangle~, \label{Umk}\\
&& \tilde V_{mk} = \langle \tilde\phi_k  |  v_m   \rangle~. \label{Vmk}
\end{eqnarray}

\item Invert $\tilde U$ and then calculate a matrix
\begin{equation}
\tilde Z~=~\tilde U^{-1}~\tilde V~.
\label{tilde Z}
\end{equation}

\item Diagonalize $\tilde Z$ to get its eigenvalues $\lambda_k$ and orthonormal
eigenmodes $\phi_k$. In the basis of $\phi_k$ the $\tilde Z$ becomes a
$Z={\rm diag}\{\lambda_1,\lambda_2,\dots\}$.

\end{itemize}
Having calculated $\lambda_k$ and $\phi_k$ we can construct the operator 
$\hat d^\dagger$ and the Bogoliubov vacuum state (\ref{Z}). Excited 
eigenstates can be obtained by repeated action of the quasiparticle creation 
operators
$\hat b_m^\dagger$ on the vacuum state.

\subsection{ Density Measurement }

  In an ideal density measurement positions of all $N$ atoms are measured.
The probability distribution for the $N$ positions in a $N$-particle quantum
state $|\Psi\rangle$ is given by
\begin{eqnarray}
&&
p_N(x_1,\dots,x_N)\sim \nonumber\\
&&
\langle\Psi|
\hat\psi^{\dagger}(x_1)\dots\hat\psi^{\dagger}(x_N)
\hat\psi(x_N)\dots\hat\psi(x_1)
|\Psi\rangle        \label{pN}
\end{eqnarray} 
We want to simulate density measurements by generating from this distribution
typical outcomes $(x_1,\dots,x_N)$. In general this is a formidable task.

  The problem becomes tractable when, for some physical reasons, it is 
possible to truncate the single particle Hilbert space to just two modes, say, 
$\phi_0(x)$ and $\phi_1(x)$ (generalization to $2,3\dots$ modes is 
straightforward). With only two modes we can adopt the 
algorithm of Javanainen and Yoo \cite{JY}. In their algorithm positions 
$(x_1,\dots,x_N)$ are not generated all at once but one after another. To begin 
with, $x_1$ is chosen randomly with a reduced single particle probability 
distribution 
\begin{equation}
p_1(x_1)\sim \langle\Psi|\hat\psi^{\dagger}(x_1)\hat\psi(x_1)
|\Psi\rangle~.
\label{p1}
\end{equation} 
Once the actual $x_1$ is chosen we calculate a state vector
\begin{equation}
|\Psi_1\rangle = 
\hat\psi(x_1) |\Psi\rangle =
\left[ \hat a_0 \phi_0(x_1) + \hat a_1 \phi_1(x_1) \right] 
|\Psi\rangle~,\end{equation}
and store it in the memory. After $(k-1)$ steps we know $(x_1,\dots,x_{k-1})$ 
and a state vector
\begin{equation}
|\Psi_{k-1}\rangle = \hat\psi(x_{k-1}) |\Psi_{k-2}\rangle~,
\end{equation}
calculated for the actual $x_{k-1}$. To generate the next coordinate $x_k$ we 
use a conditional probability distribution
\begin{eqnarray}
p(x_k &|& x_{k-1},\dots,x_1) =
\frac{p_k(x_k,\dots,x_1)}{p_{k-1}(x_{k-1},\dots,x_1)}\sim
\nonumber\\
&& 
\frac{  \langle\Psi_{k-1}| 
        \hat\psi^{\dagger}(x_k)\hat\psi(x_k) 
        |\Psi_{k-1}\rangle                     }
     {  \langle\Psi_{k-1}|\Psi_{k-1}\rangle    } ~.
\label{pcond}
\end{eqnarray}
Note that this is a function of $x_k$ only, all coordinates 
$x_{k-1},\dots,x_1$ have already been fixed. Once we get the actual outcome 
$x_k$ we calculate $|\Psi_k\rangle=\hat\psi(x_k)|\Psi_{k-1}\rangle$ and store 
it in the memory. 

  The key simplification discovered in Ref.\cite{JY} is that all single
particle probability distributions that we encounter, like
Eqs.(\ref{p1},\ref{pcond}), have the same functional form
\begin{eqnarray}
p(x) &=&  
A~|\phi_0(x)|^2 + B~|\phi_1(x)|^2 +                 \cr
&& 
C~\phi^*_0(x)\phi_1(x) + C^*~\phi_0(x)\phi^*_1(x) ~. 
\label{MarProb}
\end{eqnarray}
To find the real $A,B$ and the complex $C$ we need to sample $p(x)$ at
only four values of $x$. When $\phi_0,\phi_1$ are real, like for the dark 
soliton, then $C$ is also real, and just $3$ sampling points are enough. The 
total algorithm to generate $N$ positions $(x_1,\dots,x_N)$ scales like $N^2$, 
a slight improvement over the $N^3$ in Ref.\cite{JY}.
 
 To summarize, we provide an input $N$-particle state $|\Psi\rangle$ (in a
Fock representation build on the two single particle modes
$\phi_0,\phi_1$) and the algorithm generates a random string of
coordinates $(x_1,\dots,x_N)$ according to the $N$-particle probability
distribution (\ref{pN}). Each application of the algorithm returns a
random outcome $(x_1,\dots,x_k)$ of a single realization of density
measurement on the state $|\Psi\rangle$.

\subsection*{ Acknowledgments }

We acknowledge helpful discussions with James Anglin, Yvan Castin, 
and Jakub Zakrzewski. This research
was supported in part by KBN grants 2 P03B 092 23 (JD) and 5 P03B 088 21 (ZPK
and KS).


\end{document}